\documentclass[a4paper]{jpconf}
\usepackage{graphicx,url}
\begin{document}

\title{A tagged low-momentum kaon test-beam exposure with a 250L LAr~TPC (J-PARC T32)}

\author{J-PARC T32 collaboration:Ê~O. Araoka$^{a}$, 
A. Badertscher$^{b}$, A.~Curioni$^{b}$, S.~Di~Luise$^{b}$, U. Degunda$^{b}$, 
L. Epprecht$^{b}$, L. Esposito$^{b}$,
A.~Gendotti$^{b}$, T.~Hasegawa$^{a}$, S.~Horikawa$^{b}$, 
K. Kasami$^{a}$, N. Kimura$^{a}$, L. Knecht$^{b}$, T.~Kobayashi$^{a}$, 
C. Lazzaro$^{b}$, D. Lussi$^{b}$, M. Maki$^{a}$, A.~Marchionni$^{b}$, T. Maruyama$^{a}$,
A.~Meregaglia$^{b}$, T. Mitani$^{c}$, Y.~Nagasaka$^{c}$, 
J. Naganoma$^{c}$, H. Naito$^{d}$, S. Narita$^{d}$, G. Natterer$^{b}$, 
K.~Nishikawa$^{a}$, A. Okamoto$^{c}$, H. Okamoto$^{c}$, 
F. Petrolo$^{b}$, F. Resnati$^{b}$, A. Rubbia$^{b,}\footnote{Contact person: \mailto{andre.rubbia@cern.ch}.}$, 
C. Strabel$^{b}$, M. Tanaka$^{a}$, T.~Viant$^{b}$, Y. Yamanoi$^{a}$, 
K.~Yorita$^{c}$, M. Yoshioka$^{a}$}

\vspace*{3mm}
\address{$^{a}$  High Energy Accelerator Research Organization (KEK), Tsukuba, Ibaraki 305-0801, Japan\\
         $^{b}$ ETH Zurich, Institute for Particle Physics, CH-8093 Zurich,  Switzerland \\
         $^{c}$ Waseda University 3-4-1 Okubo, Shinjuku-ku, Tokyo 169-8555, Japan\\
         $^{d}$ Iwate University, Morioka, Iwate, 020-8551, Japan
}


\begin{abstract}
At the beginning of 2010, we presented at the J-PARC PAC an R$\&$D program towards large (100 kton scale) liquid argon TPCs, suitable to investigate, in conjunction with the J-PARC neutrino beam, the possibility of CP violation in the neutrino sector and to search for nucleon decay. As a first step we proposed a test experiment to identify and measure charged kaons, including their decays, in liquid argon. The detector, a 250L LAr  TPC, is exposed to charged kaons, in a momentum range of $540-800$ MeV/c, in the K1.1BR beamline of the J-PARC slow extraction facility. 
This is especially important to estimate efficiency and background for nucleon decay searches in the charged kaon mode ($p \rightarrow \bar{\nu} K^+$, etc.), where the kaon momentum is expected to be in the few hundred MeV/c range. A prototype setup has been exposed in the K1.1BR beamline in the fall of 2010. This paper describes the capabilities of the beamline, the construction and setting up of the detector prototype, along with some preliminary results.
\end{abstract}

\section{Introduction}
\label{sec:intro}
Liquid argon Time Projection Chambers (TPC) have been proposed and studied by the ICARUS 
Collaboration for decades (see Ref.~\cite{CITE:ICARUS} and references therein). Their efforts culminated
with the underground operation of a 600~ton-scale detector at LNGS. 
The LAr TPC provides similar track position resolution as bubble chambers, with the advantage of being continuously sensitive devices with real time readout. 
In its original design, the LAr TPC suffers from
scalability problems to the relevant mass scale for next generation experiments. 
The main challenges for the extrapolation 
are (a) very small charge signals (no amplification in liquid) (b) charge attenuation and diffusion
along the drift path (c) large wire chambers in cryogenic environment and (d) wire electric capacitance
and resistance.

Since several years, we have investigated the feasibility of a new giant next-generation and multi-purpose 
neutrino observatory based on the LAr LEM TPC concept 
with a total mass in the range of 100~kton~\cite{Rubbia:2004tz}, 
devoted to particle and astroparticle physics~\cite{GilBotella:2004bv,Cocco:2004ac,
Meregaglia:2006du,Bueno:2007um,Rubbia:2009md}.

In 2008, ETHZ and KEK groups have begun to collaborate towards the realization
of a very large 100~kton-scale detector~\cite{Badertscher:2008bp}. In synergy,  
a coherent R\&D programme is being developed under the program of IPNS and KEK Detector Technology Project and 
under the CERN Recognized Experiment (RE18) and RD51 Collaboration~\cite{RD51}.

At the 9th J-PARC PAC meeting of January 2010, we presented an R\&D path towards a proposal to search for CP violation in the leptonic sector and for proton decay, using a 100 kton scale LAr TPC on the Okinoshima island~\cite{Yoshioka:2010}, Japan, positioned along the J-PARC neutrino beam on a baseline of 658 km
(J-PARC PAC Proposal P32~\cite{Rubbia:2010PAC}).
The performance of liquid argon TPCs has been mainly studied with cosmic rays and on neutrino beams, however a test using a well-defined charged particle beam has not been performed so far. As part of our R\&D path, we proposed to expose a 250L LAr TPC to a charged particle test beam of a few hundred MeV/c momentum, enriched with kaons, at the J-PARC slow extraction facility (test experiment T32). In fact one of the most important advantages for using a liquid argon TPC detector for proton decay searches is the direct identification of the kaon track for proton decays in the kaon mode.  Our proposal was readily accepted, further attracted the interest of Iwate and Waseda University groups, and we performed a first measurement campaign with a preliminary setup in October 2010 at J-PARC.

In the two body decay $p \rightarrow \bar{\nu} K^+$, the charged kaon is emitted with  a monochromatic momentum of 340 MeV/c in case of a proton at rest, but then smeared by the nucleon Fermi motion. Such kaons are not directely visible in water Cherenkov detectors due to the high momentum threshold ($\sim 600$ MeV/c) of kaons for Cherenkov radiation in water. A positive identification of the kaon track is essential to minimize the background from atmospheric neutrino interactions. Kaons are identifiable in a LAr detector by measuring the local energy loss along the tracks as a function of the residual range, and by their decay topology~\cite{Bueno:2007um}.

\begin{figure}[h]
\begin{center}
\includegraphics[width=14cm]{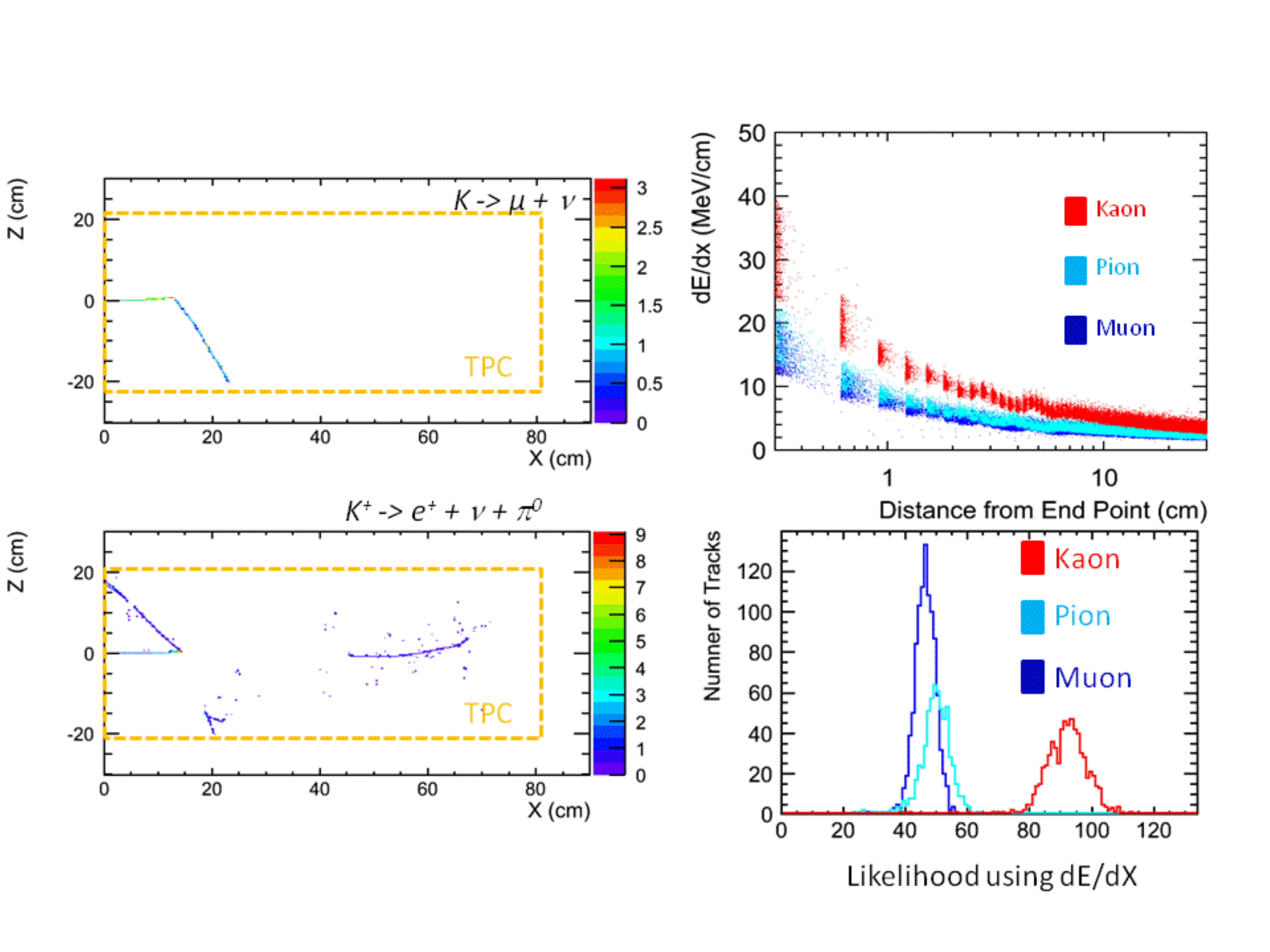}
\end{center}
\caption{\label{FIG:MC} Expected performance of the 250L prototype. Left plots; examples of kaon decays. Top-right; local dE/dx as a function of residual range, bottom-right; dE/dx based likelihood distribution for kaon identification}
\end{figure}

Fig.~\ref{FIG:MC} shows a MC simulation of kaon decays in a $40 \times 40 \times 80$ cm$^3$ LAr TPC, along with local dE/dx as a function of residual range and a dE/dx based likelihood distribution for kaons, pions and muons. 
The likelihood shows a good separation of kaons from pions and muons. In addition the topology of the decay and possibly the reconstruction of the decay products provide additional constraints on the identification of the primary particle. 
In a 100 kt LAr detector there are in total $\sim 200000$ atmospheric neutrino events/year, a fraction of which are single track events, constituting a background to the $p \rightarrow \bar{\nu} K^+$ search; this test will allow to experimentally determine the efficiency of kaon identification and the rejection factor of pions, muons and electrons.

We plan to build a 250L LAr TPC operated in double phase, where the ionization charge, extracted from the liquid, is amplified in the gas phase by means of a Large Electron Multiplier and readout by a two-dimensional projective anode, as described in~\cite{Badertscher:2008bp} and references therein.
Due to time limitations between January and October 2010, we first built a 250L prototype detector, operated in liquid phase and with coarse readout sampling. This detector has been exposed to charged particle beams, in a momentum range of 200--800 MeV/c, in the K1.1BR beamline of the J-PARC slow extraction facility. The beamline is equipped with an electro-static separator providing a $K/\pi$ ratio of $\sim 1$. Kaon identification by the LAr detector is cross checked against the identification of the incident particles provided by the beam monitor equipment in the beamline.

The beamline in the slow extraction facility, equipped with an electro-static separator, is described in section~\ref{sec:beamline}. An overview of the experimental setup, including a discussion of the beamline instrumentation, is given in section~\ref{sec:T32}.  The components and operation of the cryogenic system are described in detail in section~\ref{sec:cryosystem}, while the construction and assembly of the prototype detector are detailed in section~\ref{sec:detector}. Some preliminary results are shown in section~\ref{sec:results}.

\section{K1.1BR beamline}
\label{sec:beamline}
\begin{figure}[h]
\begin{center}
\includegraphics[width=10cm]{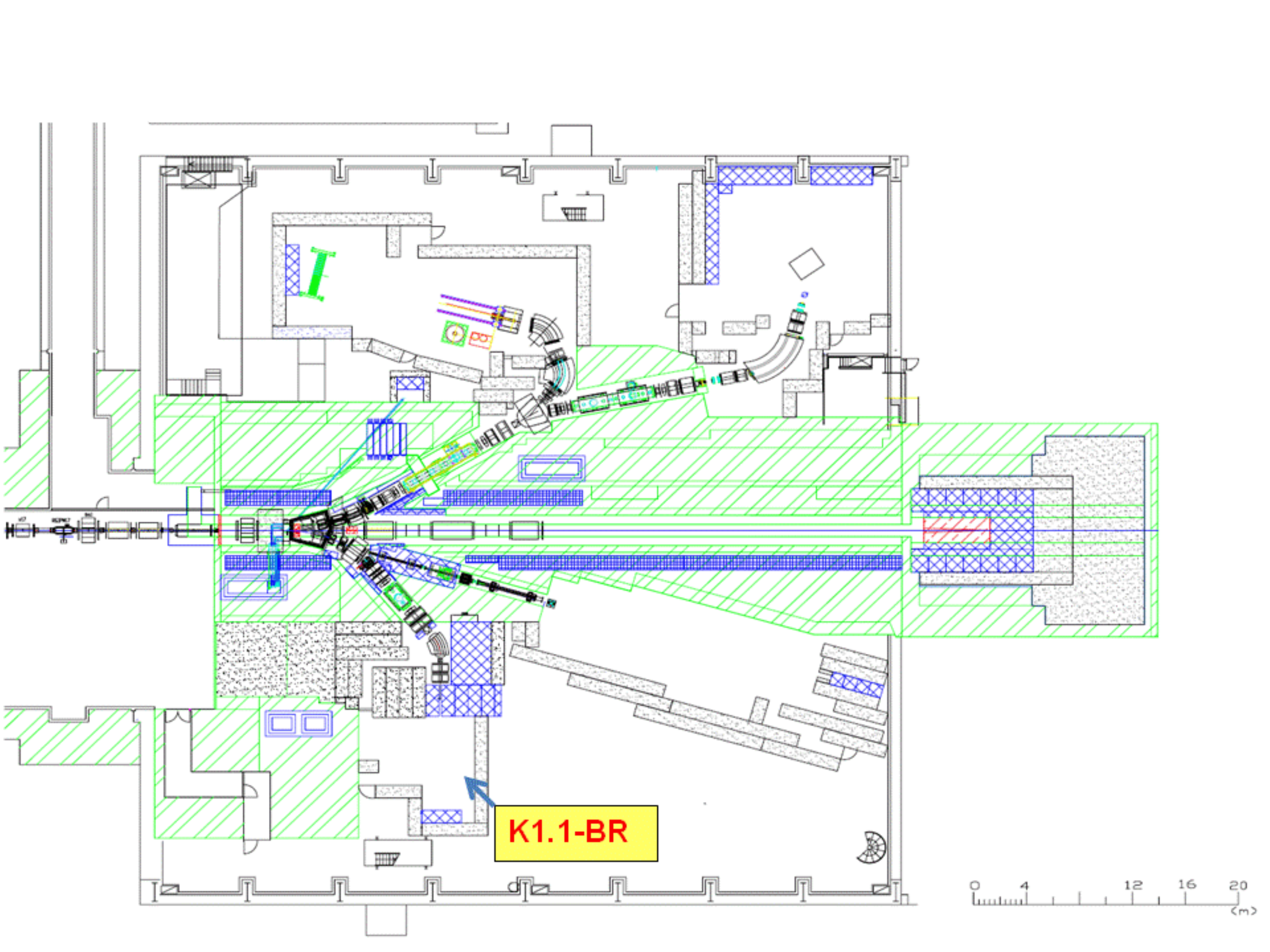}
\end{center}
\caption{\label{FIG:K11BR}J-PARC hadron hall and the location of the K1.1BR beamline}
\end{figure}

The J-PARC slow extraction facility (hadron hall) and the location of the K1.1BR beamline are shown Fig.~\ref{FIG:K11BR}. The K1.1BR beamline has been designed by the TREK Collaboration~\cite{cite:TREK}.

\begin{figure}[h]
\begin{center}
\includegraphics[width=12cm]{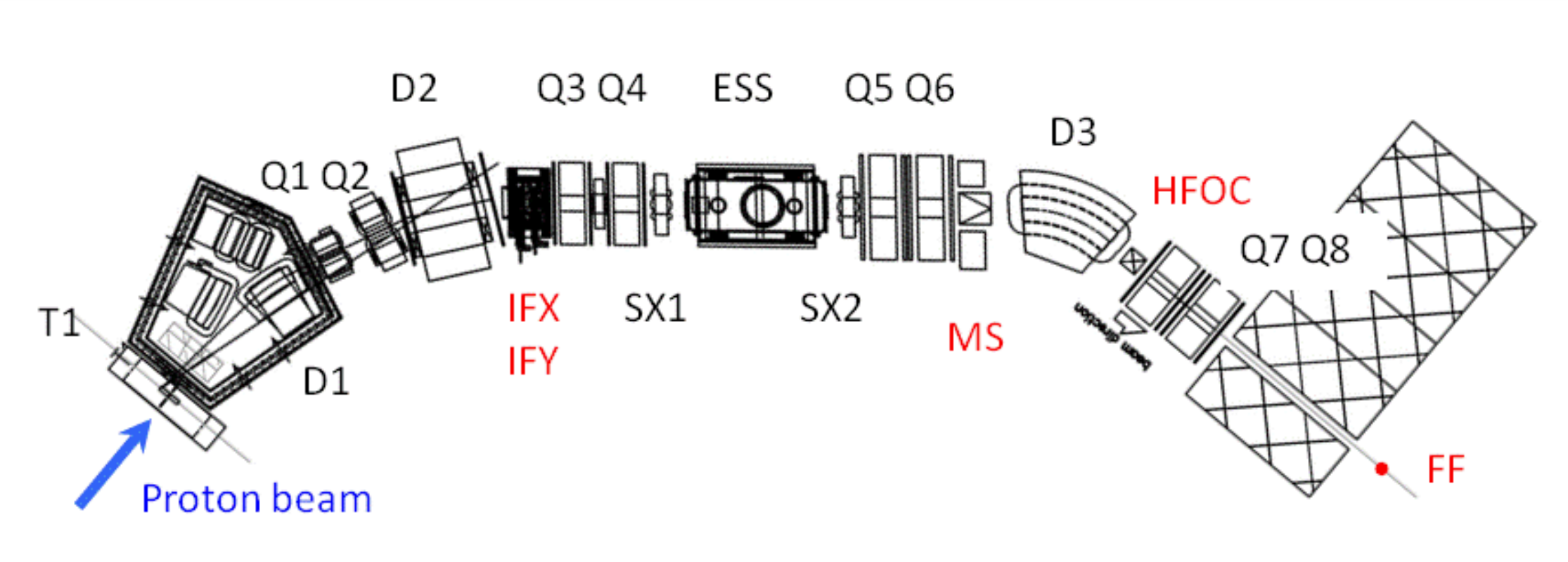}
\end{center}
\caption{\label{FIG:K11BR_2} Magnet layout of the K1.1BR beamline}
\end{figure}

The basic concept of the K1.1BR beamline is a low momentum (0.8 GeV/c) separated K$^+$ beam, by means of a single electro-static separator. The layout of the K1.1BR line, about 20 m long, is given in  Fig.~\ref{FIG:K11BR_2}.

The secondary beam takes off after the T1 production target before the first bending magnet D1. A quadrupole doublet Q1-Q2 focuses the beam at a vertical focus IFY, downstream of a second bending magnet D2. An horizontal acceptance slit is positioned at the location of the IFY vertical focus. A further quadrupole doublet Q3-Q4 makes the beam vertically parallel in the ESS electro-static separator and contains the beam horizontally. The ESS is a 2.0 m long electro-static separator, generating an electric field of 50 kV/cm over an 11 cm vertical gap. The following quadrupoles Q5 and Q6 generate a vertical focus at the mass slit (MS) position and an horizontal focus at HFOC, after the D3 bending magnet. Additionally two sextupole magnets (SX1, SX2), before and after ESS, are used to correct for higher order aberrations. The vertical slit at MS removes most of the pions, and the horizontal slit at HFOC removes 'cloud' pions (from the decay of neutral kaons close to the production target, constituting a widespread source) and slit-scattered pions. A pion rejection factor of $\sim 500$ is expected, resulting in a K/$\pi$ ratio of $\sim 1$.

The final quadrupole doublet Q7-Q8 provides the final focus, with horizontal and vertical beam rms widths of 0.8 cm and 0.5 cm, respectively. The momentum spread $\Delta$p/p is $\pm$3 $\%$. The absolute number of expected kaons is 2$\times$10$^{6}$ for 5.5$\times$10$^{13}$ 30 GeV protons on target.

It is possible to tune the K1.1BR beamline down to 200 MeV/c momentum, but 
with a kaon content sharply falling below 800 MeV/c. In order to get an acceptable kaon flux at \textless 800 MeV/c momentum, we use a degrader at the end of the beamline, where particles lose energy by ionization. We have been using a combination of lead glass blocks and a lead brick of 12.5 cm and 2.5 cm thickness, respectively.   

\section{T32 setup}
\label{sec:T32}
\subsection{Overview}
\begin{figure}[h]
\vspace{1pc}
\begin{center}
\includegraphics[width=12cm]{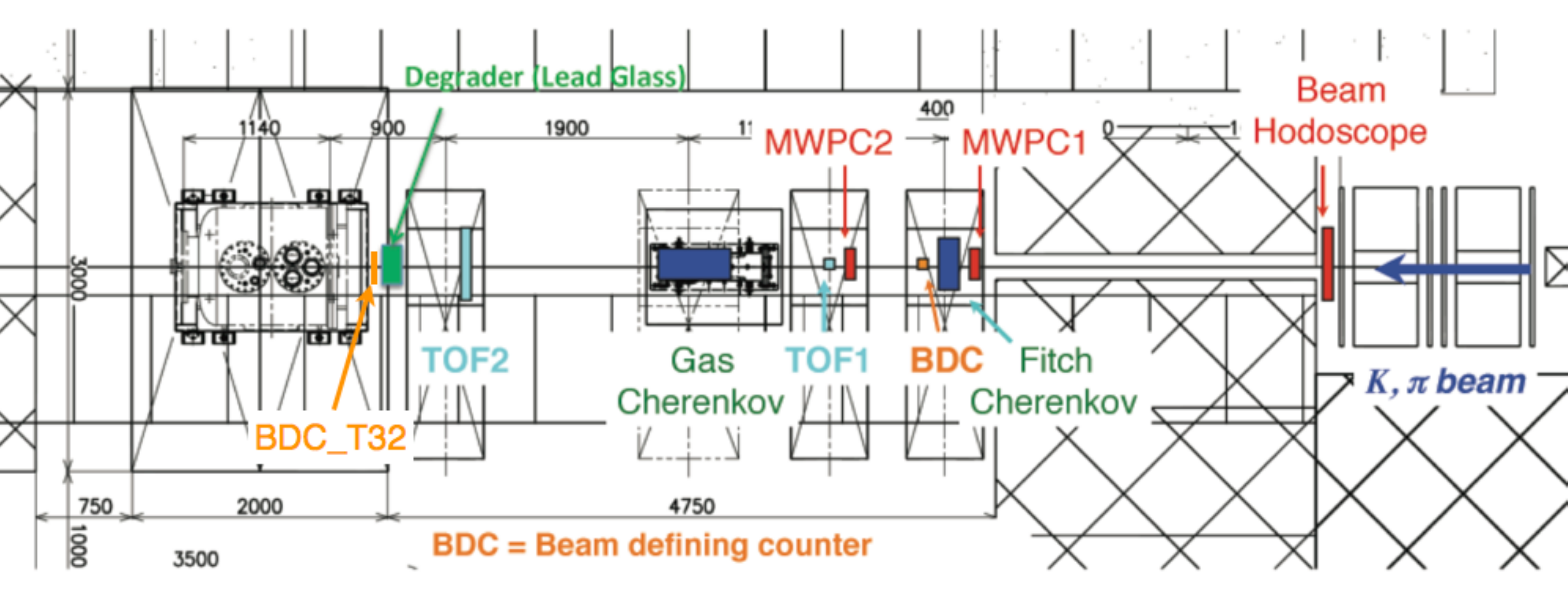}
\end{center}
\caption{\label{FIG:BE_TREK} K1.1BR beamline showing the beam monitor equipment and the location of the 250L cryostat.}
\end{figure}

Fig.~\ref{FIG:BE_TREK} shows the overview of the T32 setup. Beamline instrumentation, over a length of $\sim 5$m, is used for the definition of beam particles and for their identification. The 250L TPC chamber is situated in the downstream part of the beamline. Just upstream of the chamber it is possible to insert a degrader to decrease the momentum of the beam particles.

\subsection{Beam instrumentation}
Beam instrumentation has been provided by the TREK Collaboration. As shown in Fig.~\ref{FIG:BE_TREK}, there is a beam hodoscope, a beam defining counter (BDC), two MWPCs, two time of flight counters (TOF), a gas Cherenkov counter and a Fitch-type differential Cherenkov. In addition we have placed an additional beam defining counter (BDC\_T32) just upstream of the 250L LAr TPC and downstream of the degrader position.

\begin{figure}[h]
\begin{center}
\includegraphics[width=0.75\linewidth]{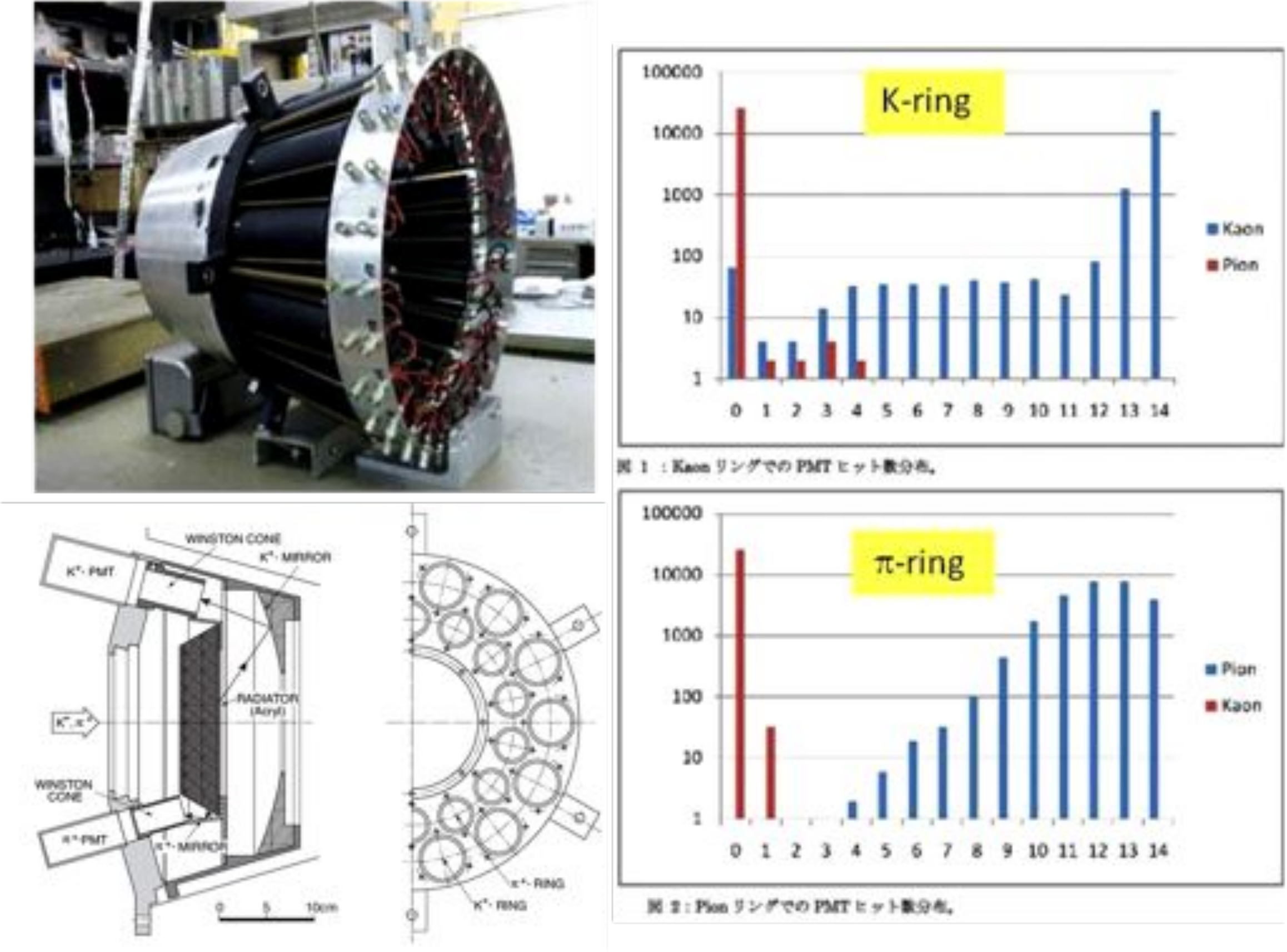}
\end{center}
\caption{\label{FIG:FC} Configuration (Left) and simulated performance (Right) of the Fitch Cherenkov counter.}
\end{figure}

In the Fitch Cherenkov counter kaons and pions go through a 40 mm-thick acrylic radiator (see Fig.~\ref{FIG:FC}--Left), where they emit Cherenkov light at definite angles depending on their velocities. The Cherenkov light from pions is reflected by a mirror around the radiator and detected by an inner PMT ring ($\pi$--ring). The light from kaons is reflected by a backward parabolic mirror and detected by an outer PMT ring (K-ring). Each ring is composed of 14 PMTs with a Winston cone at the entrance.

The PMT hit multiplicity in each ring of the Fitch counter is used to trigger online on pions or kaons. Fig.~\ref{FIG:FC}--Right shows the simulated performance. The kaon selection efficiency is $(99.5 +0.5 -0.9)\%$, 
while rejection factor of pion is close to 100$\%$. 

The gas Cherenkov counter is used in the online trigger to reject electrons and
positrons in the beam.

The TOF counters are at a distance of 3.5 m. With a measured time resolution of $\sim 200$ ps, they provide an excellent separation of kaons from pions and muon at 800 MeV/c, where the time of flight difference is $\sim 2$ ns.

\section{The cryogenic system}
\label{sec:cryosystem}
\subsection{Cryostat}
\begin{figure}[ht]
\begin{center}
\begin{minipage}{0.49\linewidth}
\includegraphics[width=0.95\linewidth]{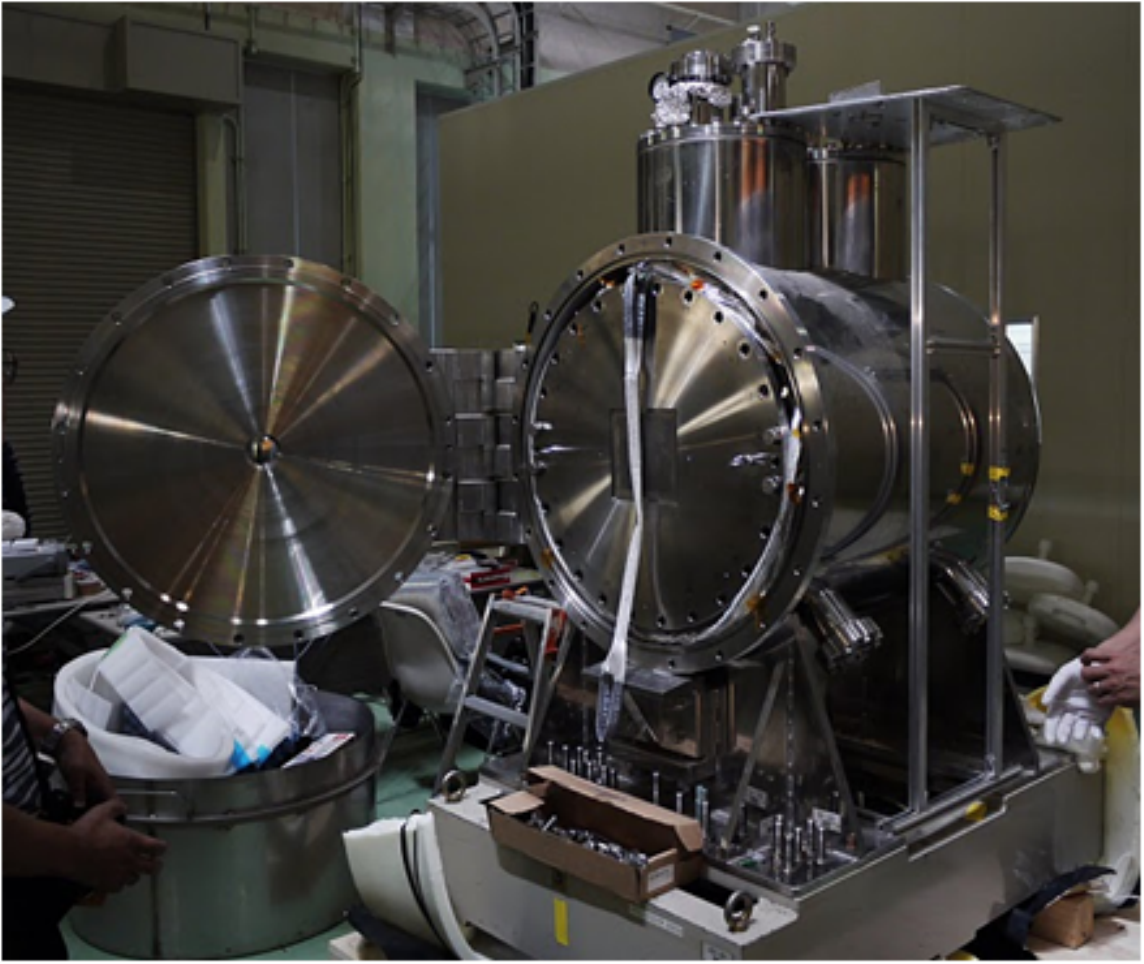}
\end{minipage}
\begin{minipage}{0.49\linewidth}
\includegraphics[width=0.93\linewidth]{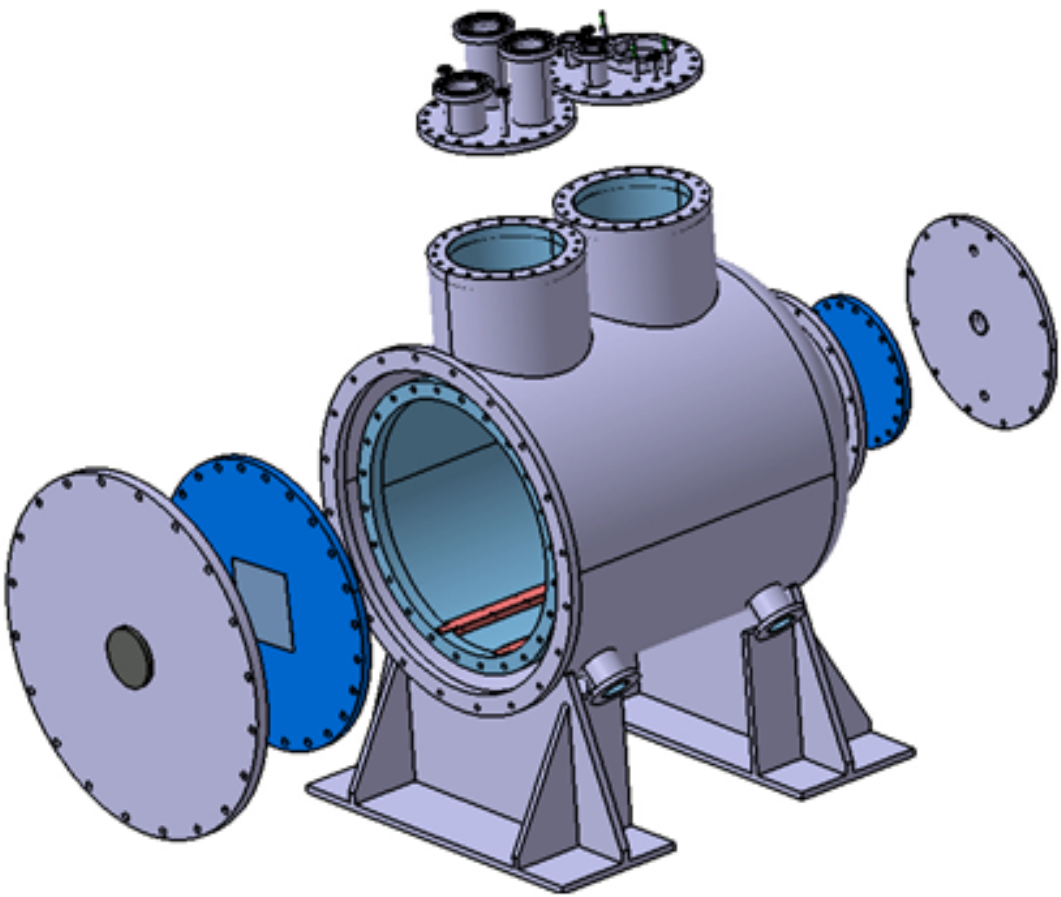}
\end{minipage}
\caption{Picture and 3D drawing of the 250L cryostat}
\label{fig:250Lcryostat}
\end{center}
\end{figure}

The cryostat of the detector has been borrowed from the MEG experiment (search for $\mu^+ \rightarrow e^+ \gamma$ conversion), and used to house a prototype LXe calorimeter exposed to gamma and electron test beams~\cite{CITE:MEG_L}. 
Fig.~\ref{fig:250Lcryostat} shows a picture and a 3D drawing of the cryostat, originally built by JECC TORISHA Co., ltd. 

It consists of two concentric vessels, an inner and an outer one. The inner vessel, of about 75 cm diameter and 100 cm length, is filled with LAr. The maximum size of a parallelepiped which could fit inside the inner vessel is ~$50 \times 50 \times 100$ cm$^3$ (250L).
The volume between the two vessels is evacuable and additionally it is thermally insulated with super-insulating material (20 layers of aluminized mylar film). The inner vessel is supported by four cylinders made of GFRP (Glass Fiber Reinforced Polymer) to reduce the heat inflow. The resulting heat load is only $\sim$30W at LAr temperature, as measured by monitoring the evaporation of LAr.

There are 4 flanges to access the inner volume. One of the two top flanges is devoted to cryogenics, and it houses the cryoocoler input, the LAr and LN$_2$ inlets, and the safety devices. On the other top flange, there are signals and high voltage feedthroughs for the detector and the slow control instrumentation. The front flange is used to insert the TPC detector into the vessel. A 9 cm diameter beam window is present at the center of the front flange, and it is equivalent to a material thickness of 0.13 X$_{0}$. The back flange is connected to a vacuum pump for vacuum insulation.

\subsection{Cooling and purification system}
\begin{figure}[h]
\begin{minipage}{0.42\linewidth}
\includegraphics[width=0.95\linewidth]{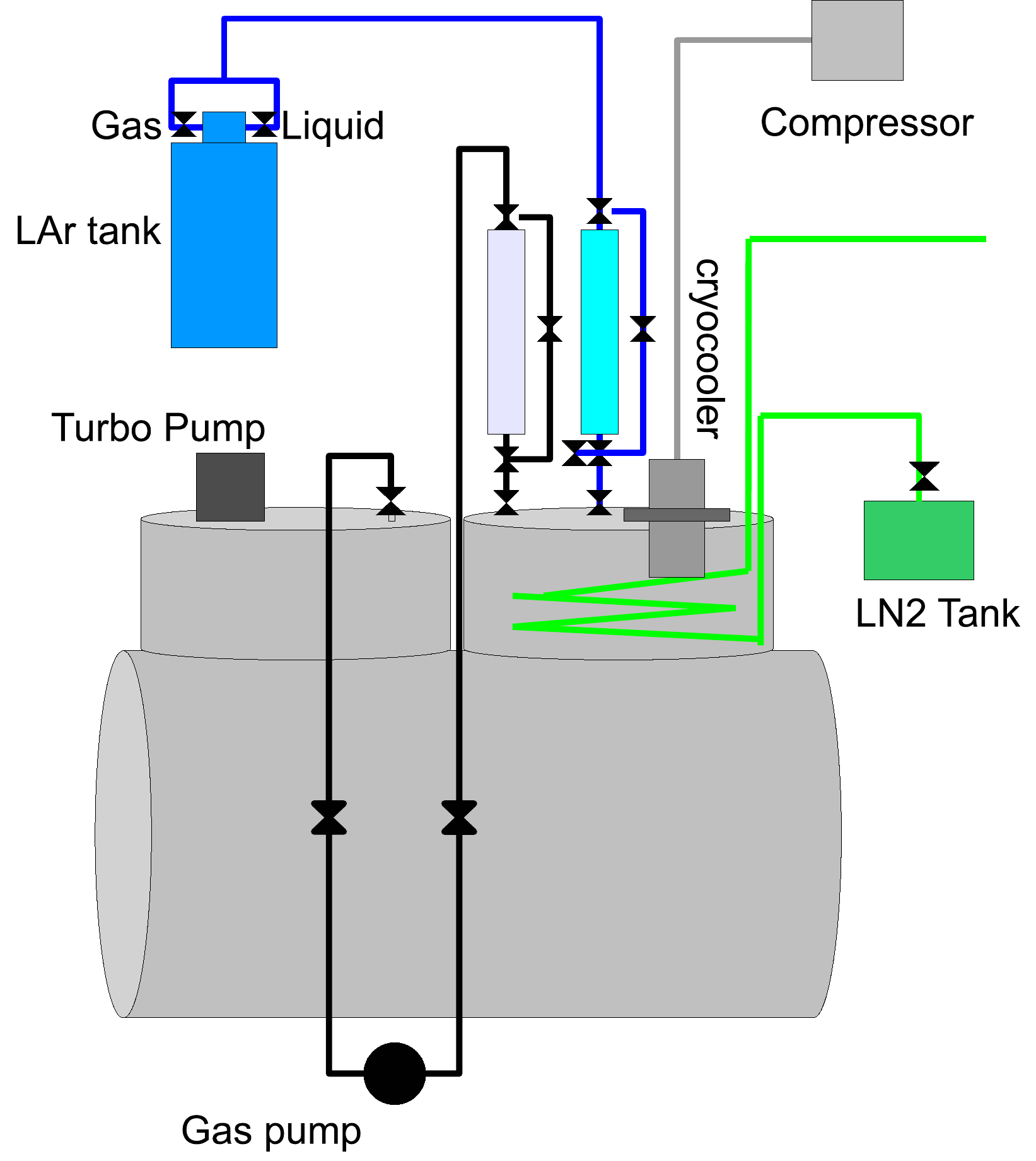}
\end{minipage}
\begin{minipage}{0.22\linewidth}
\includegraphics[width=0.95\linewidth]{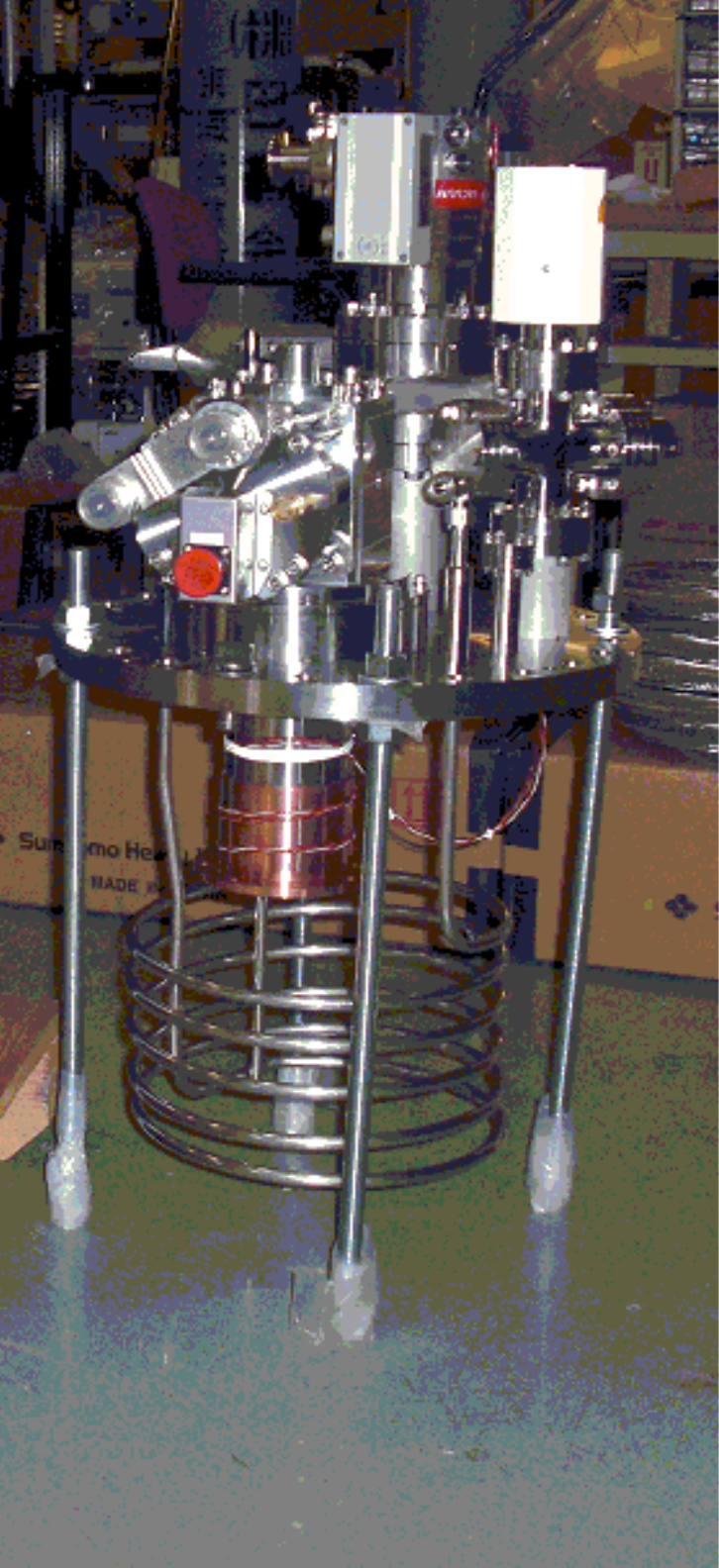}
\end{minipage}
\begin{minipage}{0.35\linewidth}
\includegraphics[width=0.95\linewidth]{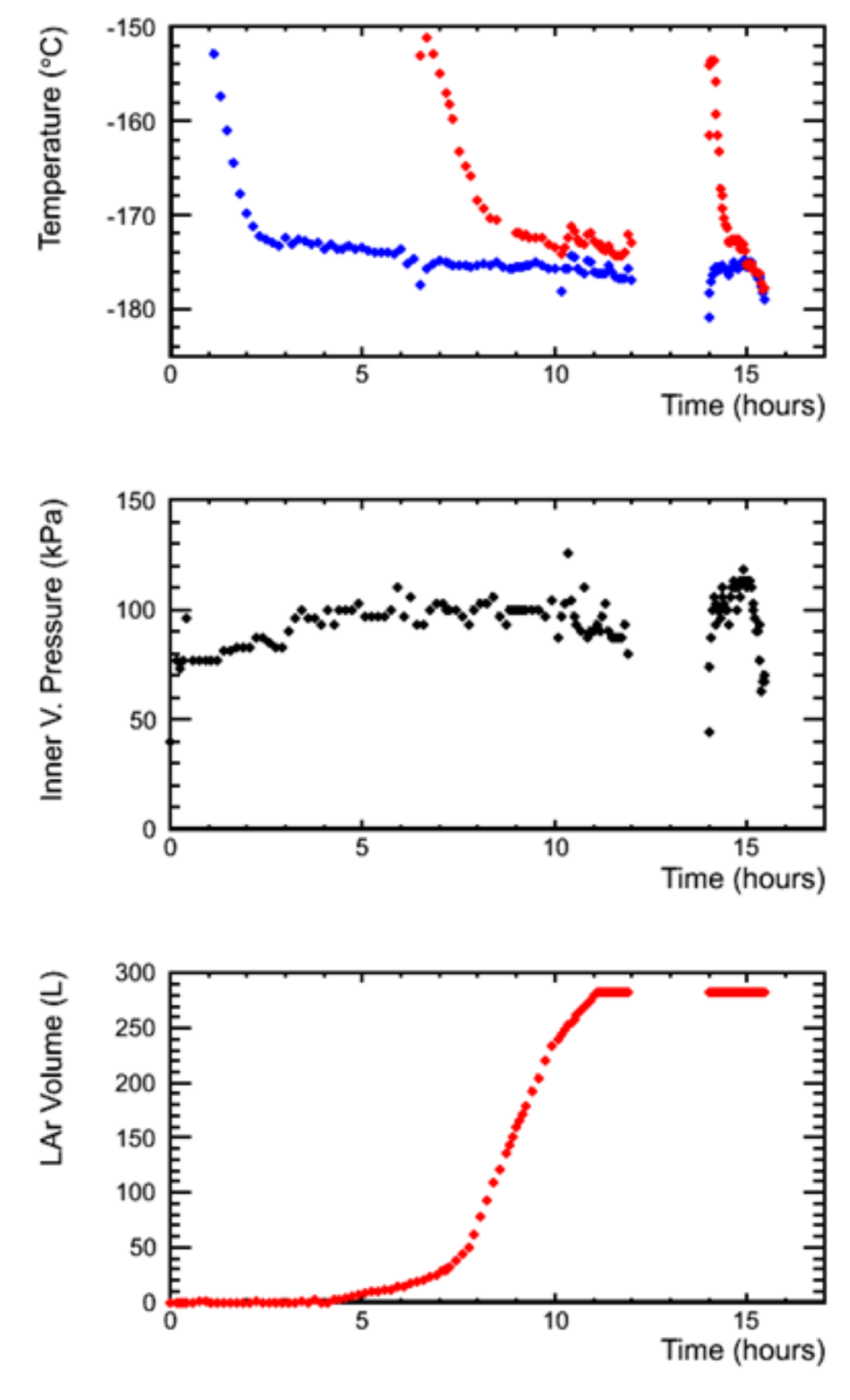}
\end{minipage}
\caption{\label{FIG:CryoSetup} Left: schematic of the cryogenic setup and purification system. Middle: a picture of the GM cryocooler (the copper cold head is visible at the bottom) and of the LN$_2$ coil. Right: monitoring of the system while filling with LAr. Top plot shows the temperature of the cryocooler head (blue) and of the detector anode (red). Middle plot shows the inner vessel temperature, bottom plot the LAr volume.} 
\end{figure}

A schematic of the cryogenic setup and purification system is shown in Fig.~\ref{FIG:CryoSetup}--Left. 

A Gifford-MacMahon(GM) cryocooler and a LN$_2$ coil inside the inner vessel provide the necessary cooling. They are mounted on one of the top flanges, as shown in Fig.~\ref{FIG:CryoSetup}--Middle.
The cryocooler cold head has $\sim$ 100 mm diameter, with a length of 193 mm inside the inner vessel. The available cooling power at LAr temperature is 160 W. In stable operating conditions, after filling and in absence of gas recirculation, a heater on the cold head is used to prevent the argon from freezing, since the cryocooler power is larger than the heat load.
The LN$_2$ coil is located on the same top flange housing the cryocooler. The coil is made of 5 turns of 1 mm thick 3/8" stainless steel pipe, for a total length of $\sim 400$ cm.  The coil provides a cooling power of more than 600 W for a LN$_2$ flow of 20 L/hour.   

In order to achieve a good argon purity, commercial LAr during the initial filling is passed through a purification cartridge. Additionally a gas recirculation and purification system has been implemented as shown in Fig.~\ref{FIG:CryoSetup}--Left.
We use a custom-made purification cartridge~\cite{CITE:MASA} for the liquid argon filling phase, with a diameter of 60 mm and a length of 600 mm, containing activated copper-coated alumina granules~\cite{CITE:CuO} to remove O$_{2}$ and type 4A molecular sieve material to remove H$_{2}$O, as previously used in~\cite{CITE:ARGONEUT}. The two components are kept in separate volumes in the cartridge, with sintered metal disks at the inlet and outlet of the cartridge for the containment of the granules. The cartridge contains 1125 g and 363 g of copper-coated granules and molecular sieve material, respectively, corresponding to a volume ratio of 2:1.

In the recirculation system, argon gas from the inner vessel is forced by a metal bellows pump to circulate through a gas purification filter and then inserted back into the inner vessel. The piping is made of 3/8" stainless steel tubes, using VCR connections and welds to minimize leaks. The gas purification filter, a SAES~\cite{CITE:SAES} type MC3000, provides at the output $<$0.1 ppb of oxygen and water, at a maximum gas flow rate of about 80 slpm.

\subsection{Monitoring, control and safety}
\begin{figure}[htbp]
\begin{center}
\includegraphics[width=10cm]{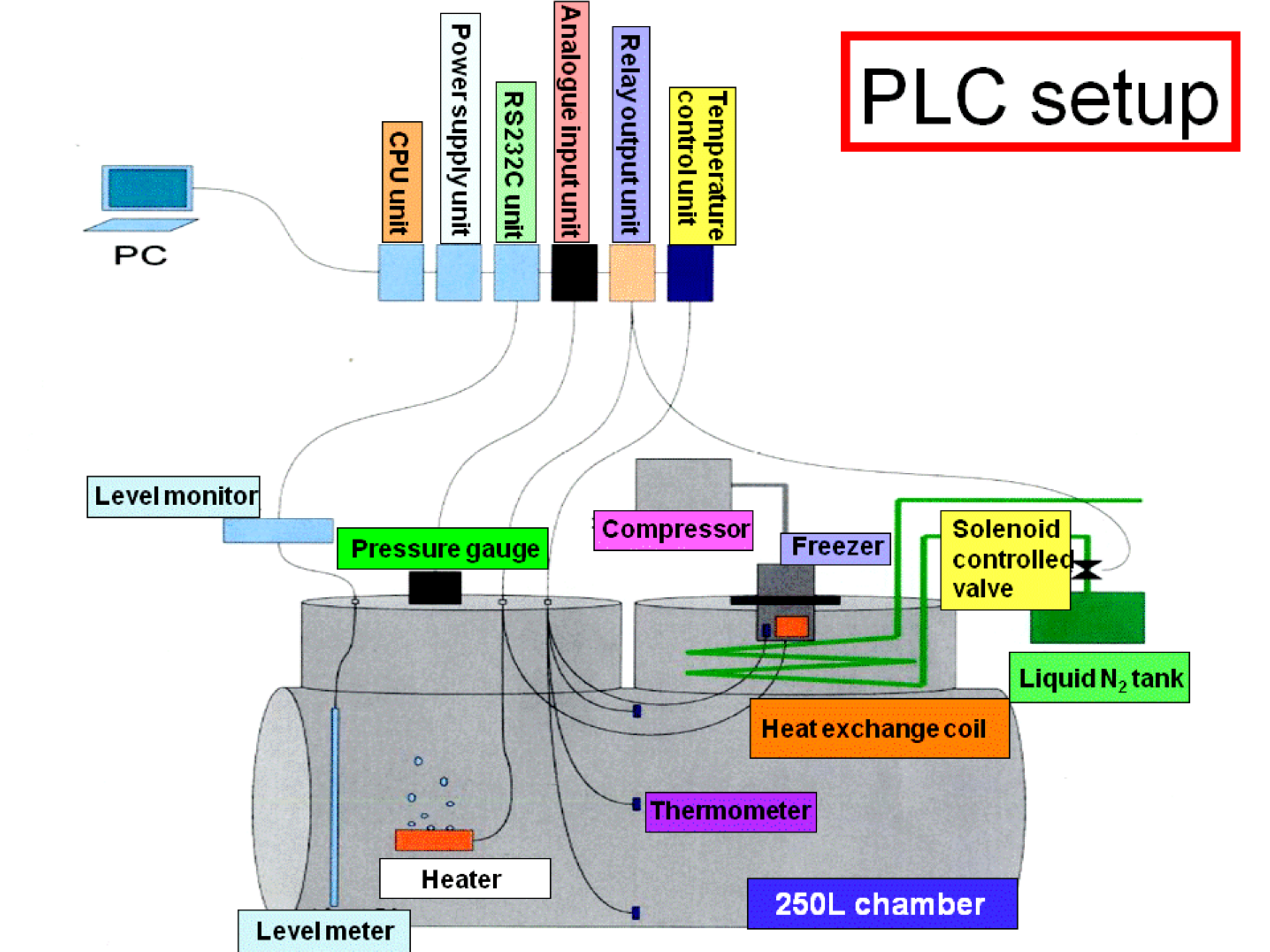}
\end{center}
\caption{\label{FIG:PLC} Setup of the PLC monitoring and control.}
\end{figure}

The monitoring and control of the cryogenic system is managed by a PLC from Keyence~\cite{CITE:KEYENCE}.
The detailed setup is shown in Fig.~\ref{FIG:PLC}.

The following items are monitored and controlled by the PLC:
\begin{itemize}
\item temperatures using PT100 sensors;
\item pressure in the inner vessel. If the  pressure exceeds a set value, LN$_2$ is let to flow in the cooling coil;
\item heater on the cryocooler cold head. This is controlled by the PLC in order to keep the temperature in the vessel at a given set value.
\end{itemize}

As the inner vessel contains a large amount of cryogenic liquid in a closed volume, it is necessary to protect it with the aid of pressure relief devices. The safety of the system is assured by a purge line equipped with a rupture disk, a safety valve, and an electro-valve. The safety valve opens at 0.15 MPa (Gauge), while the burst pressure of the rupture disk is 0.2 MPa (Gauge). The electro-valve is opened in case of electric power outage. Additionally two oxygen monitors are installed in the K1.1BR beamline area for personnel protection.

\subsection{250L Filling Procedure}
The process of filling the inner vessel with LAr requires particular attention not to exceed the maximum allowable pressure in the vessel and not to compromise the purity of the argon because of outgassing during the transition phase to LAr temperature. We describe here in detail each step performed:
\begin{itemize}
\item Evacuation of the inner vessel using a molecular turbo pump (Edwards, 300L/s capacity) and a getter pump (SAES, 40 L/s capacity), directly mounted on the top flange for $\sim$ 1 week.
We achieved a vacuum level of 1$\times$10$^{-4}$ Pa, with an outgasssing rate of $\sim 1$ Pa/hour.

\item Filling of the inner vessel with argon gas, passing through the purification cartridge, up to 80--100 kPa(Gauge).

\item Initial cooling of the vessel using the GM cryocooler and the LN$_2$ heat exchanger coil. 

\item Gas recirculation is turned on at the same time, with a typical gas flow of 70 L/min.

\item As temperature and pressure inside the vessel decrease,
we insert argon gas and we keep the pressure in the inner vessel at $\sim 80$ kPa(Gauge).

\item After $\sim$5 hours, the vessel is cooled down to LAr temperature, so that LAr starts accumulating at the bottom of the vessel. The typical liquefaction speed, limited by the cooling power available in the cryocooler and in the LN$_2$ coil, is 10 L/hour.

\item At this point, we start filling the vessel directly with LAr, going through the purification cartridge. It takes $\sim$3 hours to cool down the purification cartridge to LAr temperature. The filling speed, limited by the impedance of the purification cartridge, is $\sim$ 50 L/hour.

\item Filling is stopped once the readout anode is immersed in LAr.

\item We recirculate the argon gas, evaporated from the liquid inside the vessel, in order to improve or maintain the initial purity.
\end{itemize}

Figure~\ref{FIG:CryoSetup}--Right shows the monitoring values while filling the inner vessel with LAr. The top plot shows the temperature of the cryocooler head (blue) and of the detector anode (red), the middle plot shows the inner vessel
temperature and the bottom plot the volume of LAr from the level meter measurement. Note that the filling was suspended at $\sim$12 hours and resumed at $\sim$14 hours; the LAr volume measurement saturates at 280 L because of a level meter limitation.

\section{The LAr TPC}
\label{sec:detector}
\subsection{TPC detector}
For the first test campaign in October 2010, we built a prototype detector with coarse readout sampling, operated in liquid argon phase. A set of PMTs, at the bottom of the TPC, is used to get a trigger for cosmic ray muons from the argon scintillation light.
Table~\ref{TAB:250LPara} summarizes the main detector parameters. 

\begin{table}[h]
\begin{center}
\begin{tabular}{cc} 
\hline \hline 
Total LAr volume & $\sim$ 300 L \\
Field cage dimension & 42 cm $\times$ 42 cm $\times$ 78 cm \\
Fiducial volume & 40 cm $\times$ 40 cm $\times$ 76 cm \\
Fiducial mass & 170 kg \\
Electric Field & 0.2--0.3 kV/cm \\
Maximum cathode voltage & 12 kV \\
Readout method & charge collection in liquid phase \\
Readout pitch & 1.0 cm \\
Number of readout channels & 76 \\
\hline
\end{tabular}
\caption{
	Parameters of the 250L prototype detector exposed at the J-PARC K1.1BR beam in October 2010.}
\label{TAB:250LPara}
\end{center}
\end{table}

\begin{figure}[h]
\begin{center}
\includegraphics[width=0.80\linewidth]{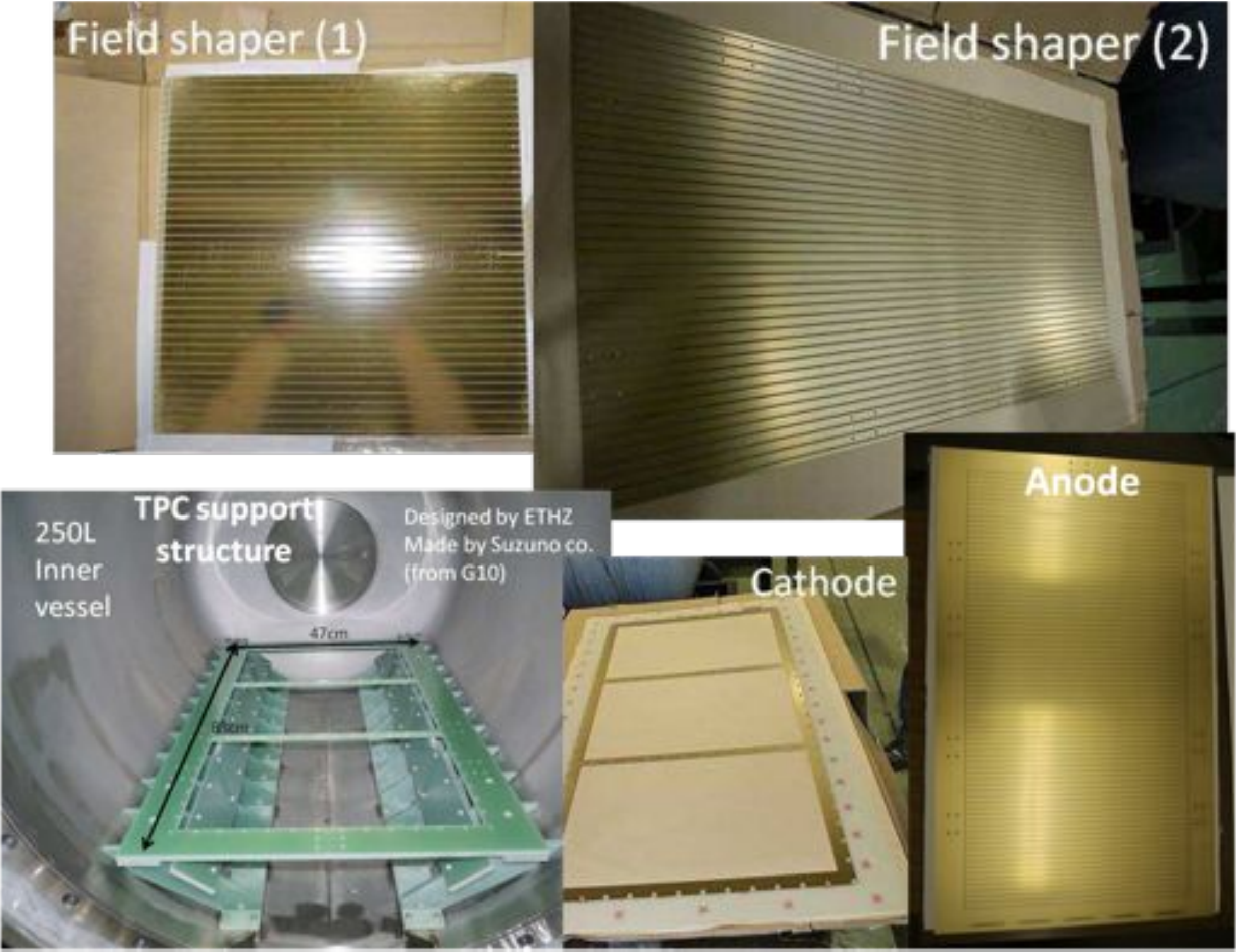}
\end{center}
\caption{\label{FIG:FCT} Detector components and support structure}
\end{figure}

The field cage determines the fiducial volume of the detector. Its dimensions are $42 \times 42 \times 78$ cm$^3$, for a total LAr fiducial mass of 170 kg. It is made with a printed circuit board (PCB) technique, where the field shapers are 8 mm wide gold-plated copper strips, with 10 mm pitch, on a 1.6 mm thick PCB (see top pictures in Fig.~\ref{FIG:FCT}). The copper strips are connected to a voltage divider, made of a chain of 10 M$\Omega$ high voltage resistors.
The cathode plane (see Fig.~\ref{FIG:FCT}), located at the bottom of the field cage, needs to be transparent to the argon scintillation light generated in the fiducial volume, detected by the PMTs at the bottom. The cathode is built as a grid, with 100 $\mu$m stainless steel wires, spaced by 5 mm. 
Electric field simulations of this configuration show a good uniformity of the drift electric field, with distortions of $\sim 10 \%$ at 1~cm distance from the field cage walls.

\begin{figure}[h]
\begin{center}
\includegraphics[width=0.80\linewidth]{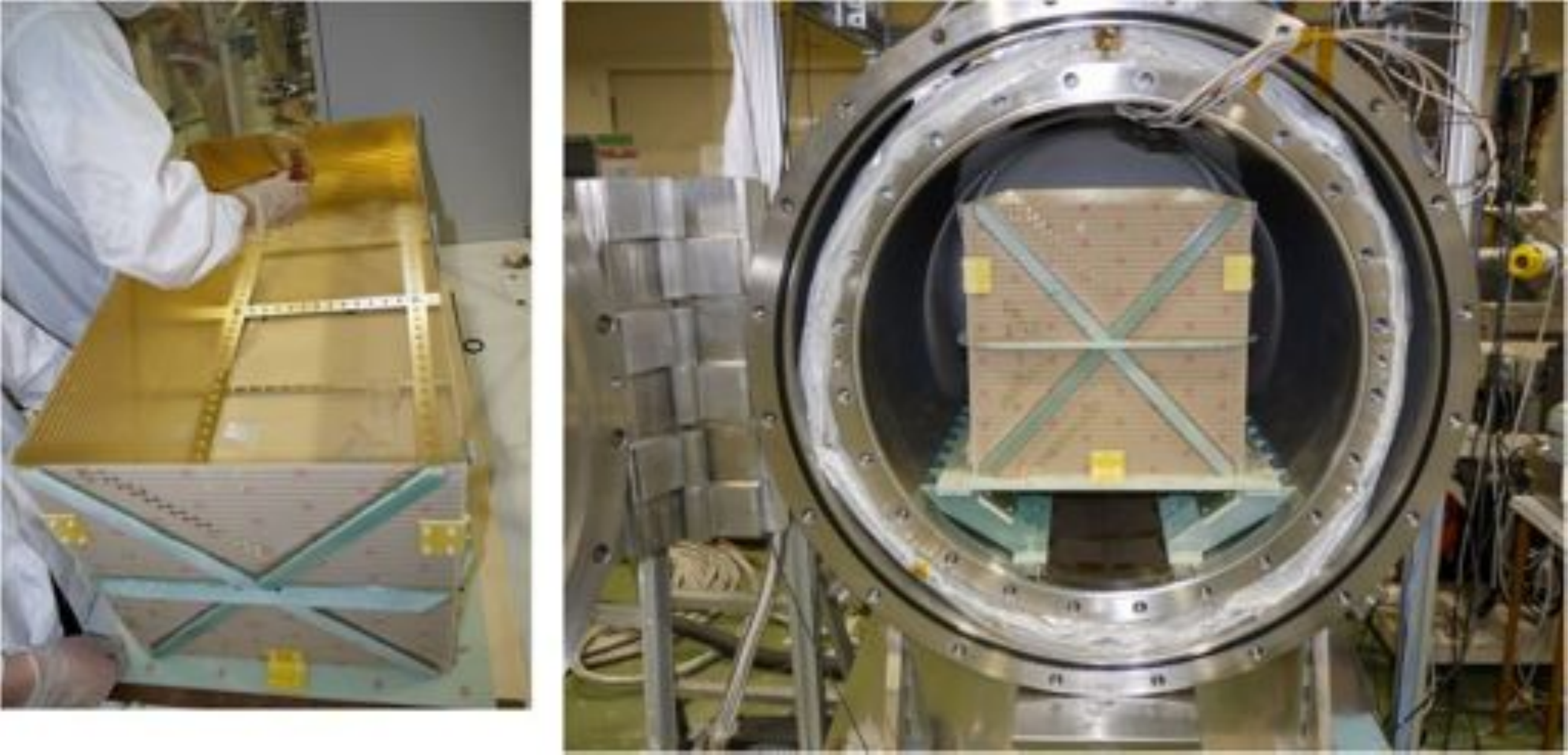}
\end{center}
\caption{\label{FIG:FC_AA} TPC detector during assembly (left) and after insertion inside the inner vessel on the support structure at the bottom (right).}
\end{figure}

Ionization electrons are drifted upwards to a collecting anode, immersed in LAr, segmented in 1 cm wide strips for a total of 76 readout channels (see Fig.~\ref{FIG:FCT}). Fig.~\ref{FIG:FC_AA} shows the assembled detector inside the inner vessel.

\begin{figure}[h]
\begin{minipage}{0.6\linewidth}
\includegraphics[width=0.7\linewidth]{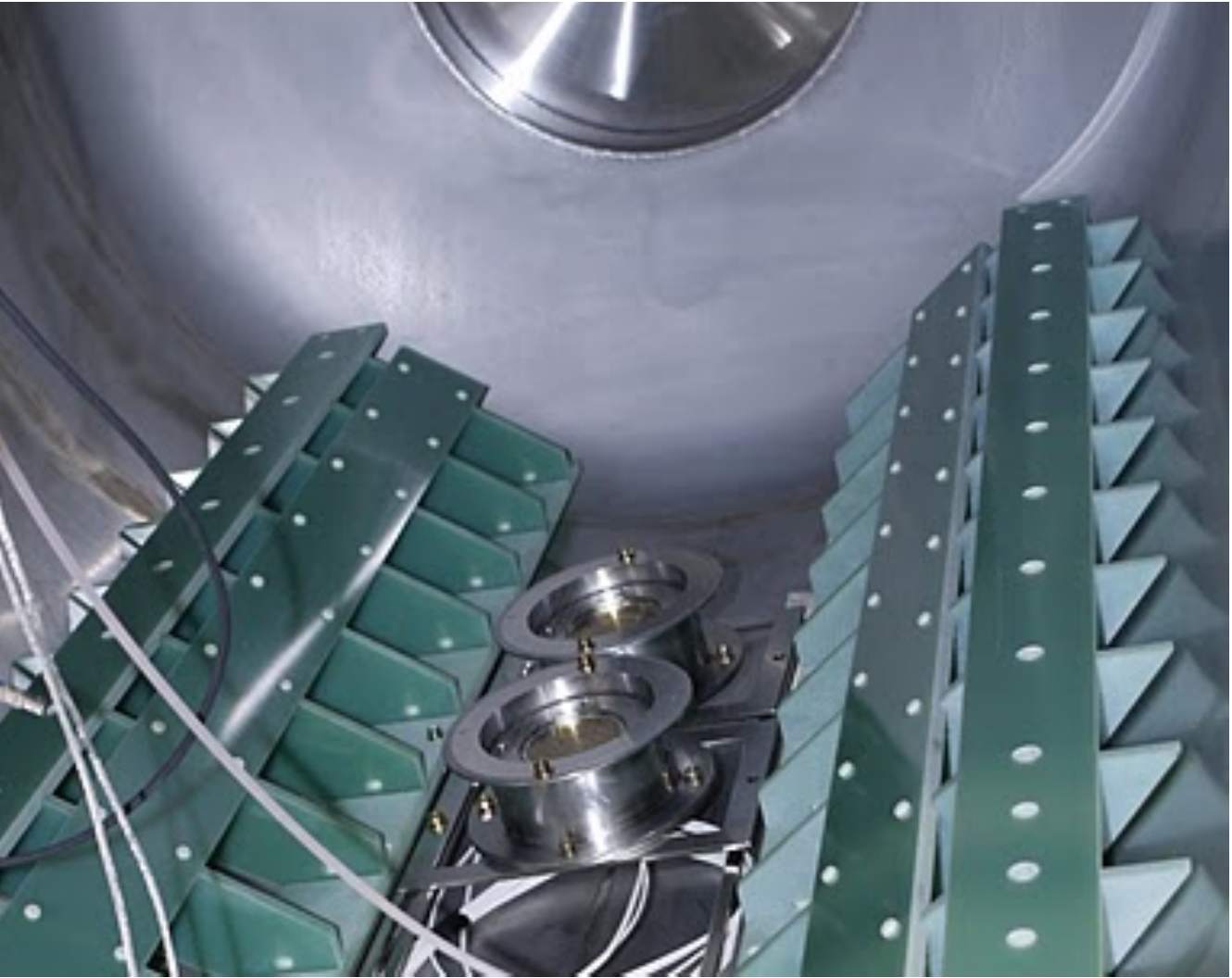}
\caption{\label{FIG:PMTloc} PMTs with stainless steel holders at the bottom of the vessel.}
\end{minipage}\hspace{2pc}%
\begin{minipage}{0.30\linewidth}
\includegraphics[width=0.60\linewidth]{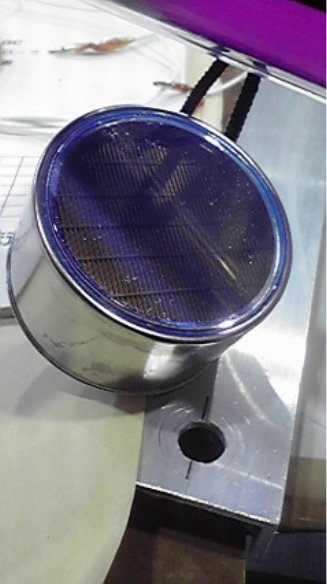}
\caption{\label{FIG:PMTdip} TPB coated PMT, illuminated by UV light.}
\end{minipage} 
\end{figure}

Two cryogenic PMTs are located below the cathode, as shown in Fig.~\ref{FIG:PMTloc}. They are Hamamatsu type R6041-02ASYM MOD, operated at $\sim 800$ V with a typical gain of 1.0$\times$10$^{6}$. 
The PMTs are supported by stainless holders with protection grids on top. The grids are made of a stainless steel crossed wire pattern of 100 $\mu$m width and 10 mm pitch, obtained from etching, and they are kept at ground voltage to protect the PMTs from the high voltage values of the nearby cathode plane.  

In order to shift the 128 nm scintillation light from LAr to a sensitive region of the PMTs,we use tetraphenyl-butadiene (TPB) as wavelength shifter, which has an emission spectrum between 400 and 480 nm. The PMTs are coated with TPB, embedded in a polymer matrix. We used a mixture of TPB (0.10 g), Paraloid (1.01 g) and Toluene (19.99 g). Fig.~\ref{FIG:PMTdip} shows a TPB coated PMT illuminated by UV light.

\subsection{Trigger, electronics and DAQ}
Different triggers have been used during our data taking period, combined with different settings of the K1.1BR beamline, as summarized here:
\begin{itemize}
\item a cosmic ray trigger, by triggering on the coincidence of the two PMTs at the bottom of the 250L TPC, which detect the LAr scintillation light produced in the fiducial volume;
\item a basic beam trigger, formed by the coincidence of the Beam Defining Counter BDC, the two time of flight counters TOF1 and TOF2 and the BDC\_T32 counter placed just upstream of the 250L chamber (see Fig.~\ref{FIG:BE_TREK});
\item a positron trigger, from the coincidence of the beam trigger with the signal from the gas Cherenkov;
\item a kaon trigger, from the coincidence of the beam trigger with the signal from the Fitch Cherenkov.
\end{itemize}

The collected charge on the segmented anode of the 250L LAr chamber is readout by the SY2791 system~\cite{CITE:CAEN} from CAEN, developed in collaboration with ETHZ.
Charge sensitive preamplifiers, with a sensitivity of $~\sim 12$ mV/fC, are directly mounted on the CAEN modules. The ouput signals are digitized by 12 bit ADCs with 2.5 MHz sampling frequency.  

\begin{figure}[htbp]
\begin{center}
\includegraphics[width=0.6\linewidth]{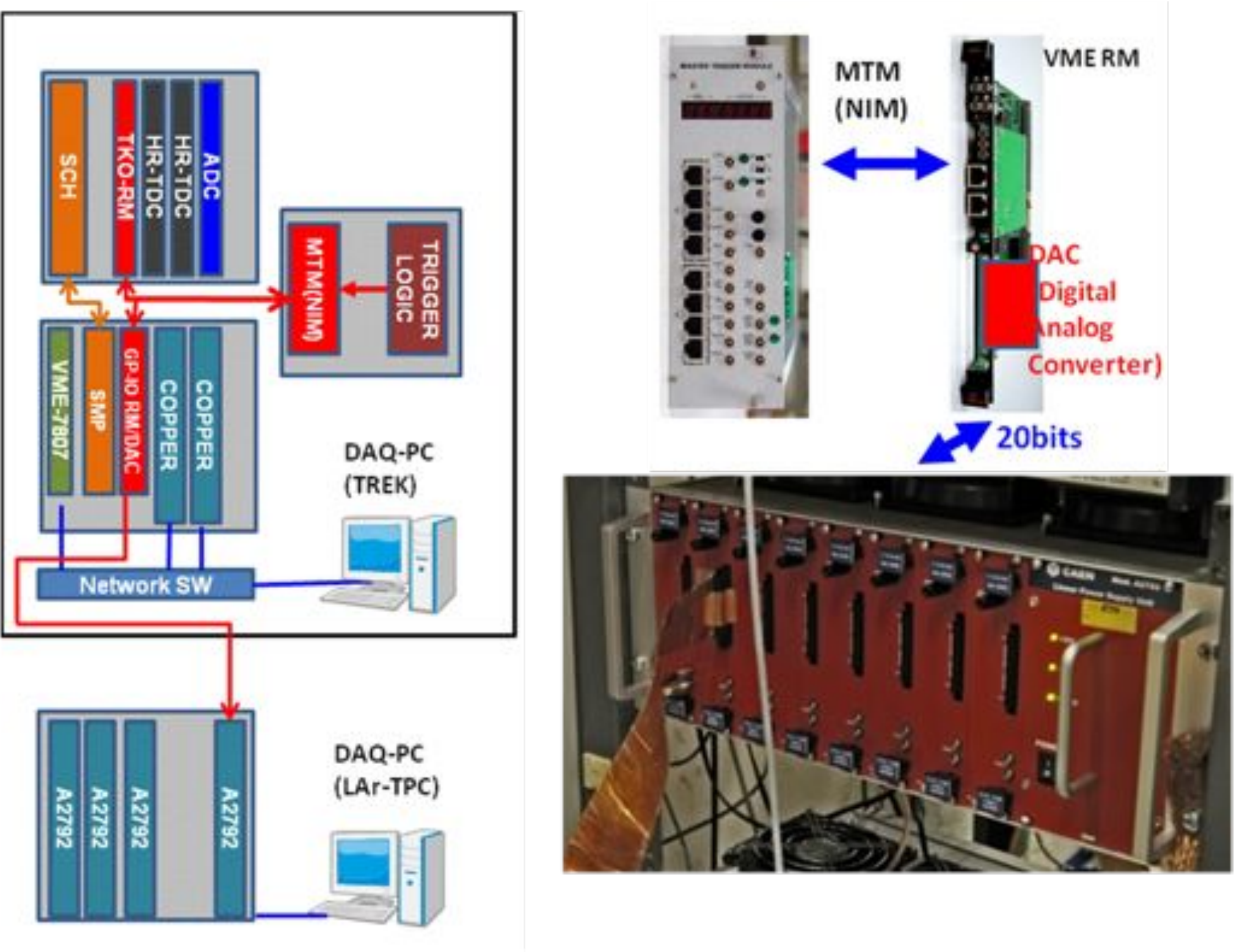}
\end{center}
\caption{\label{FIG:MTM}Event tag system between beamline information and LAr TPC. The SY2791 system from CAEN is shown at the bottom on the right}
\end{figure}

In order to synchronize the event number between the LAr TPC detector and the beamline instrumentation we use a Master Trigger Module (MTM). The MTM module delivers a spill and a trigger number via VME (see Fig.~\ref{FIG:MTM} right) to the LAr TPC electronics and to the beamline instrumentation. Fig.~\ref{FIG:MTM} shows the configuration of the system. Data from the LAr TPC and the beamline instrumentation are merged at a later stage by using the MTM information.

\section{Preliminary results and future plans}
\label{sec:results}
After filling in October 2010, the setup has been stably operating for a week, corresponding to the allocated time at the J-PARC K1.1BR hadron hall. 
The detector has been exposed to charged particle beams in the $200-800$ MeV/c momentum range. 

\begin{figure}[h]
\begin{center}
\includegraphics[width=0.95\linewidth]{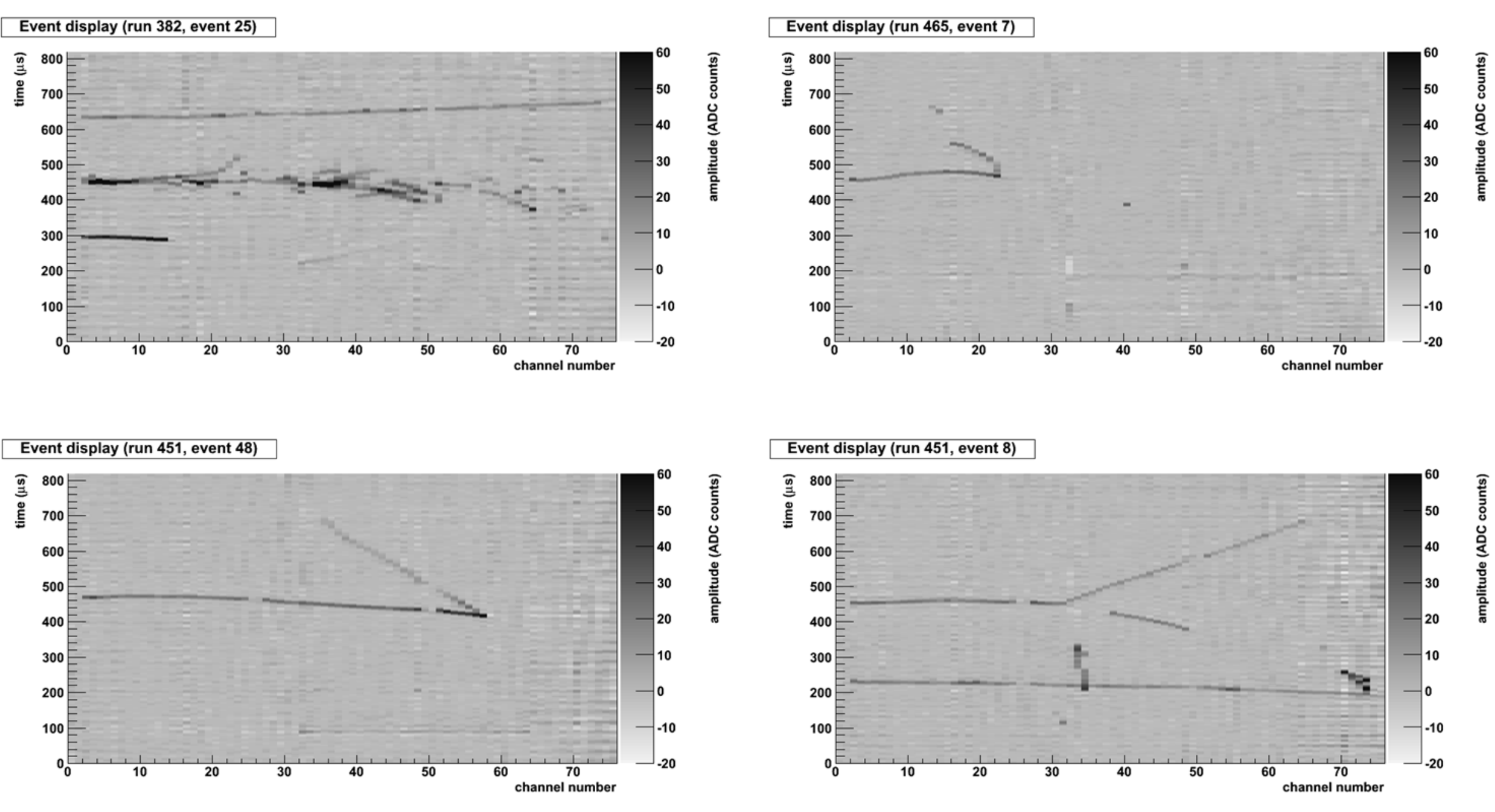}
\end{center}
\caption{\label{FIG:250Levents} Displays of collected events, after removal of coherent noise. Drift time is shown vs channel number, grey scale is proportional to signal amplitude. 
}
\end{figure}

Some images of collected events are shown in Fig.~\ref{FIG:250Levents}. The event displays show drift time vs. channel number, where channel '0' is the upstream channel and the grey scale is proportional to the signal amplitude. A digital filter has been applied to the signals to remove the coherent noise common to all channels. 

The top left picture shows an event with three overlapping beam particles crossing the detector, interpreted, from top to bottom, as a pion, a positron and a proton.
The event was acquired by triggering on positrons, with the beam momentum set to 800 MeV/c and with no degraders inserted.
The top right picture shows a candidate for a pion decay at rest, obtained by triggering on beam particles, with positron rejection, in a 200 MeV/c beam. 
The events at the bottom in Fig.~\ref{FIG:250Levents} have been taken with a kaon trigger at 800 MeV/c momentum. A lead glass block and a lead brick were inserted just before the 250~L chamber to degrade the momentum of beam particles entering the chamber.
Bottom left picture is a candidates for $K^+ \rightarrow \mu^+ \nu_\mu$ decay at rest. The bottom right picture can be interpreted as a $K^{+}_{\mu 3}$ decay in flight.

The analysis of the data and comparison to a detailed MC simulation of the setup and beam properties, is presently ongoing. 
By the observation of cosmic muons, we measured an electron lifetime in LAr better than $300 \mu s$, corresponding to an upper limit of $\sim 1$ ppb of electronegative impurities.

At the same time we have been proceeding with the construction of a double phase LAr TPC, equipped with a Large Electron Multiplier and a two-dimensional projective anode with 3 mm readout sampling, for a total of 512 channels. All components have been manufactured, and we are presently assembling them. A test with cosmic rays has been planned
at CERN. If successful, we will consider and proposed another installation in the K1.1BR beamline.

\section{Acknowledgments}
We would like to warmly thank the TREK collaboration, and in particular Jun Imazato, for information on the K1.1BR beamline and for the use of the beamline instrumentation. We also thank the MEG collaboration for lending us the cryostat for an extended period of time. We especially thank Satoshi Mihara for his valuable information on the cryostat. We greatly appreciate the help received by the J-PARC Hadron Group for the installation of our setup. 
We also thank the RD51 Collaboration for useful discussions and help of the CERN TS/DEM group.
We acknowledge the financial and technical support of our funding agencies and in particular KEK IPNS, KEK Detector Technology Project, the Swiss National Science Foundation and ETH Zurich.

\end{document}